\documentclass[preprint,aps,showpacs,floats,
superscriptaddress,nofootinbib]{revtex4}  
\usepackage{amsmath}
\usepackage{amssymb}
\usepackage{graphicx, psfig}
\newcommand{\NLSP}{\tilde{\chi}_1^0}

\newcommand{\TeV}{{\rm TeV}}
\newcommand{\GeV}{{\rm GeV}}
\begin{document}

\preprint{YITP-03-63}
\preprint{UT-ICEPP 03-07}
\title{Study of the Gauge Mediation Signal 
with Non-pointing Photons
at the CERN LHC}
\author{Kiyotomo Kawagoe}
\affiliation{Department of Physics,
Kobe University, Kobe 657-8501, Japan}
\author{Tomio Kobayashi}
\affiliation{ICEPP,
University of Tokyo, Tokyo 113-0033, Japan}
\author{Mihoko M. Nojiri}
\affiliation{YITP, Kyoto University, Kyoto 606-8502, Japan}
\author{Atsuhiko Ochi}
\affiliation{Department of Physics,
Kobe University, Kobe 657-8501, Japan}
\date{\today}

\begin{abstract}
In this paper we study the gauge mediation signal with the ATLAS 
detector at the CERN LHC. We focus on the case where the NLSP is the
long-lived lightest neutralino ($\tilde{\chi}^0_1$) which decays
dominantly into a photon ($\gamma$) and a gravitino ($\tilde{G}$).  
A non-pointing photon from the neutralino decay can be detected 
with good position and time resolutions by the electormagnetic 
calorimeter (ECAL), while the photon momentum would be precisely
measured if the photon is converted inside the inner tracking detector before
reaching the ECAL. A new technique is developed to determine the
masses of the slepton ($\tilde{\ell}$) and the neutralino from events
with a lepton and a converted non-pointing photon arising from the
cascade decay $\tilde{\ell}\rightarrow \ell\tilde{\chi}^0_1\rightarrow
\ell\gamma \tilde{G}$.  A Monte Carlo simulation at a sample point
shows that the masses would be measured with an error of 3\% for
$\cal{O}$(100) selected $\ell\gamma$ pairs.  Once the sparticle masses
are determined by this method, the decay time and momentum of the
neutralino are solved using the ECAL data and the lepton momentum
only, for all $\ell\gamma$ pairs without the photon conversion.  We
estimate the sensitivity to the neutralino lifetime for $c\tau=10$~cm
to $\cal{O}$(10)~m.
\end{abstract}


\maketitle
     
\section{Introduction}

Minimal Supersymmetric Standard Model (MSSM) is a promising candidate
beyond the Standard Model.  As the supersymmetry (SUSY) must be
spontaneously broken, the MSSM needs an additional sector (the hidden
sector) which breaks the supersymmetry while avoiding the FCNC
problem.  The origin of the SUSY breaking and the mediation to the
MSSM sector are therefore the key feature of SUSY models.  The hidden
sector SUSY breaking are expressed in terms of the order parameter of
the SUSY breaking $F$ and the scale of the SUSY breaking mediation to
the MSSM sector $M$.  The mass scale of MSSM sparticles $M_{SUSY}$ is
then of the order of $\lambda F/M$, where $\lambda$ is the coupling of
the hidden sector to the MSSM sector.  If $M\sim M_{pl}$,
$M_{SUSY}=1$~TeV corresponds to $\sqrt{F}\sim10^{10}$~GeV.  This class
of mediation is called the supergravity (SUGRA) model.  On the other
hand, the SUSY breaking mediation may be due to renormalizable
interactions, such as the gauge interaction.  This is called ``gauge
mediation'' (GM) models~\cite{Dine:1995ag,Giudice:1998bp}.  In the GM
models $M$ and $F$ are arbitrary and we expect $M\ll M_{\rm pl}$.

The GM models are described by a few parameters.  The MSSM gaugino
masses $M_i$ $(i=1,2,3)$ and slepton masses are of the order of
$\alpha_i F/M$ in the simplest GM model, where $\alpha_i$ denotes each
gauge coupling constant.  On the other hand, the gravitino mass
$m_{\tilde{G}}$ is proportional to $F_0/M_{\rm pl}$ where $F_0$ is the
order of the SUSY breaking of the total system
($F_0>F$)~\cite{Volkov:jd,Deser:uq}.

Because $M\ll M_{\rm pl}$, the lightest SUSY particle (LSP) is the
gravitino ($\tilde{G}$) in the GM models.  The next lightest SUSY
particle (NLSP) is a particle in the MSSM sector which decays into a
gravitino~\cite{Fayet:vd}.  If the lightest neutralino
($\tilde{\chi}^0_1$) is the NLSP, the dominant decay mode is
$\tilde{\chi}^0_1\rightarrow \gamma \tilde{G}$.  The neutralino
lifetime $c\tau$\footnote{In this paper the neutralino lifetime is
multiplied by the light velocity $c$ to have a dimension of length.}
is a function of $F_0$ and $m_{\tilde{\chi}^0_1}$, and may be
long-lived.

The CERN large hadron collider (LHC) is a $pp$ collider at center of
mass energy of 14~TeV.  The LHC is expected to start its physics runs
in 2007.  The initial integrated luminosity will be 10~fb$^{-1}$/yr at
the beginning (low luminosity runs), and then upgrade to
100~fb$^{-1}$/yr (high luminosity runs).  Signatures of the GM models
at the LHC are spectacular~\cite{Ambrosanio:1996jn,
Dimopoulos:1996yq,Bagger:1996bt}.  In the case of the neutralino NLSP,
SUSY events have nearly always hard photons, which may not be pointing
to the interaction point (non-pointing photons).

In this paper we propose a new approach to study the signature of the
GM models using the ATLAS detector at the LHC for the case where the
neutralino NLSP dominantly decays into a photon and a gravitino with
$c\tau$ longer than $\cal{O}$(10)~cm.  We use two newly developed
techniques.  One is to determine the direction of the gravitino
momentum by using the arrival position, arrival time, and momentum of
the non-pointing photon, which are measured at the electromagnetic
calorimeter (ECAL).  The precision of the photon momentum would be
significantly improved if the non-pointing photon is converted into an
$e^+e^-$ pair in the inner tracking detector located inside the ECAL.
The other technique is the ``mass relation method'', a mass
reconstruction technique which does not rely on the conventional
endpoint measurement~\cite{TDR}.  In this method we use the fact that
each event from a same cascade decay satisfies mass shell conditions
of intermediate particles.  These techniques are described in Sec.~II
and a simulation is carried out in Sec.~III.  A fast Monte Carlo
simulation shows that, using the new techniques, {\it statistical
error} of the masses of $\tilde{\ell}$ and $\tilde{\chi}^0_1$ can be a
few \% for ${\cal O}$(100) selected $\ell\gamma$ pairs from the decay
chain $\tilde{\ell}\rightarrow
\ell\tilde{\chi}_1^0 \rightarrow 
\ell\gamma \tilde{G}$.
Although more works, especially full detector simulations,  
are needed to establish the techniques,
the result of the fast simulation is quite encouraging. 

In Sec.~IV we show that events with $\ell\gamma$ pairs are fully
reconstructed by using the measured mass and the ECAL information.
This measurement is utilized to determine the neutralino lifetime for
$10$ cm $<c\tau< \cal O$(10)~m.  This analysis does not require the
photon to be converted in the inner detector.  Therefore available
number of events is significantly larger.  Finally in Sec.~V we
discuss how these measurements would be translated into the
fundamental parameters $F_0$, $F$ and $M$ in the GM models.

\section{\boldmath Kinematics of the events 
with non-prompt $\tilde{\chi}^0_1$ decay}

\begin{figure}
\centerline{
\centerline{\psfig{file=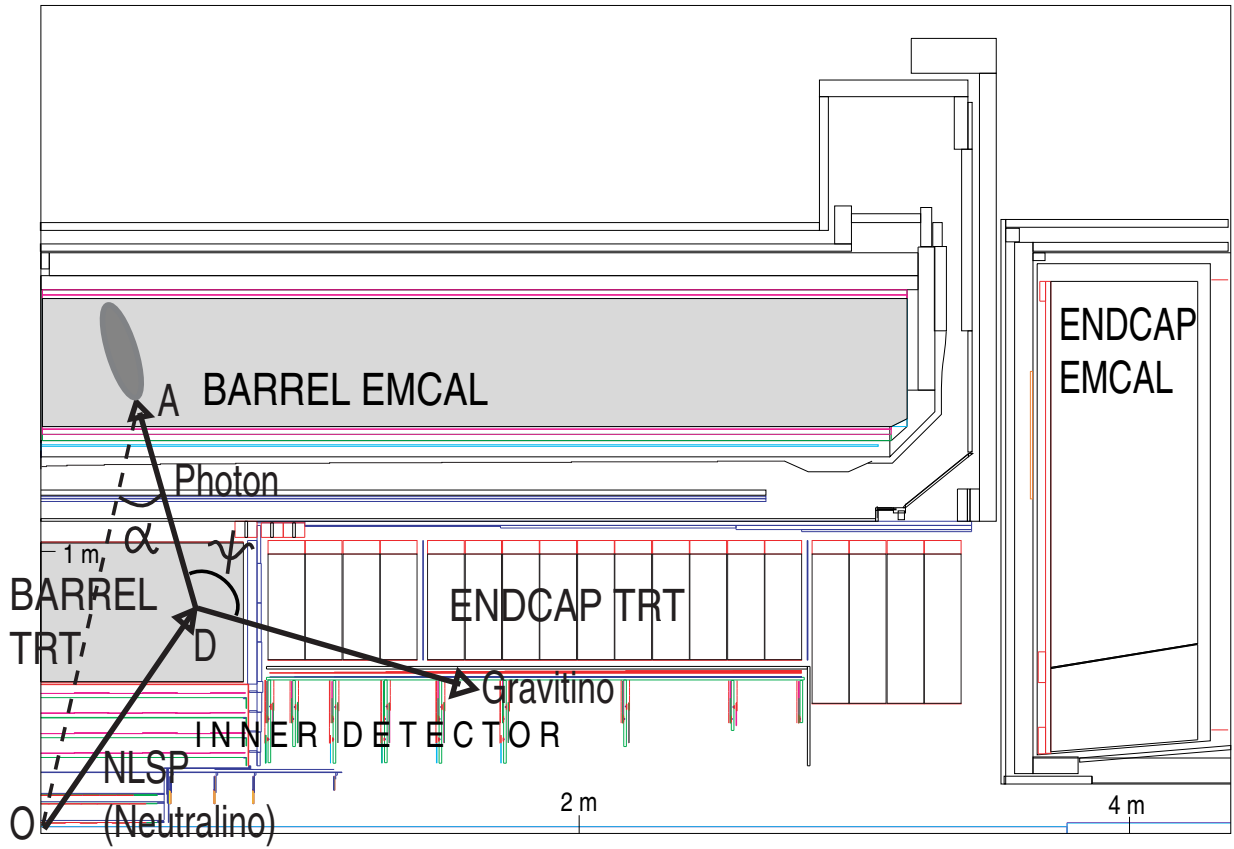,width=\columnwidth}}
}
\caption{\label{nlsp}Decay kinematics of the NLSP 
(the lightest neutralino  $\NLSP$) in the ATLAS detector.
}
\end{figure}

We first discuss the decay kinematics of the neutralino with an
unnegligible lifetime.  In FIG.~\ref{nlsp} we schematically show a
neutralino decaying into a gravitino and a photon in the ATLAS
detector.  The neutralino is produced at the interaction point O at
the time $t=0$, and then flies to the decay point D, where the
neutralino decays at $t=t_D$ into a photon and a gravitino.  The
photon goes to the point A in the ECAL at $t=t_{\gamma}$.  We define
two angles $\alpha$ and $\psi$, the former between the photon momentum
$\vec{p}_{\gamma}$ and the position vector
$\vec{x}_{\gamma}=\overrightarrow{OA}$, and the latter between
$\vec{p}_{\gamma}$ and the gravitino momentum $\vec{p}_{\tilde{G}}$.

We can experimentally measure $\alpha$, $t_{\gamma}$ and the distance
$L=\vert\overrightarrow{OA}\vert$.  The angle $\psi$ can then be
calculated from the three observable as
\begin{eqnarray}
\cos\psi &=& \frac{1-\xi^2}{1+\xi^2}\ , \cr
{\rm where }\ \ \xi  &\equiv& \frac{c t_{\gamma}-L\cos\alpha}{L\sin\alpha}.
\label{kobayashi}
\end{eqnarray}
Because  the three momenta $\vec{p}_{\tilde{\chi}}$, 
$\vec{p}_{\gamma}$ and   
$\vec{p}_{\tilde{G}}$ are on a same  plane, 
the  direction of the gravitino momentum  is completely 
determined\footnote{The formula can be shown quite easily by going 
into the  frame where the interaction point and 
the detection point is same.  We define the 
rapidity of the gravitino and photon, taking 
the boost direction as the $z$ direction. 
To obtain the formula Eq.~(\protect\ref{kobayashi})
one then goes back to the laboratory frame 
noting the additivity of the rapidity. We thank Dr. Odagiri 
for pointing out this. }.

The information of the gravitino direction may be used to determine
the sparticle masses. We describe this idea for the decay chain
$\tilde{\ell}\rightarrow \ell\tilde{\chi}^0_1\rightarrow \ell\gamma
\tilde{G}$. The slepton $\tilde{\ell}$ may be 
copiously produced at the LHC from $\tilde{\chi}^0_2$ or
$\tilde{\chi}^{\pm}_1$ decays, where $\tilde{\chi}^0_2$ and
$\tilde{\chi}^{\pm}_1$ are dominantly produced from gluino or squark
decays.  The neutralino and slepton masses ($m_{\tilde{\chi}_1^0}$ and
$m_{\tilde{\ell}}$) are related by the following formula;
\begin{eqnarray}
m_{\tilde{\ell}}^2 & = & (p_{\gamma} + p_{\tilde{G}} +p_{\ell})^2 \nonumber \\
   & = & 2 E_{\gamma} E_{\tilde{G}} (1-\cos\psi) 
     +   2 E_{\ell}E_{\tilde{G}}  (1-\cos\theta_{\ell\tilde{G}}) \nonumber \\ 
   & + & 2 E_{\ell} E_{\gamma} (1-\cos\theta_{\ell\gamma}) \nonumber \\
   & = & \left( 1 + \frac{E_{\ell} (1-\cos\theta_{\ell\tilde{G}})}
                          {E_{\gamma} (1-\cos\psi)} \right) m_{\NLSP}^2 
        + 2 E_{\ell} E_{\gamma} (1-\cos\theta_{\ell\gamma}) 
      \nonumber \\
   & \equiv & a m_{\NLSP}^2 + b\ ,
\label{mass}
\end{eqnarray}
where we use the relation 
\begin{eqnarray}
m_{\NLSP}^2 & = & (p_{\gamma} + p_{\tilde{G}})^2 \nonumber \\
          & = & 2 E_{\gamma} E_{\tilde{G}} (1-\cos\psi)\ ,
\end{eqnarray}
and neglect the lepton and gravitino masses.  A pair of parameters
$(a,b)$ can be calculated event by event from the momenta of the
lepton and the photon, and the direction of the gravitino momentum.
Because the sparticle masses $m_{\NLSP}$ and $m_{\tilde{\ell}}$ should
be common for all events, one can determine them if we have at least
two tagged events.

In Refs.~\cite{TDR,Hinchliffe:1998ys} the masses
$m_{\tilde{\chi}^0_1}$ and $m_{\tilde{\ell}}$ are determined by
measuring endpoints of mass distributions for events containing
$\ell\ell\gamma$ which come from the decay chain
$\tilde{\chi}^0_2\rightarrow
\ell\tilde{\ell}\rightarrow 
\ell\ell\tilde{\chi}^0_1 
\rightarrow 
\ell\ell\gamma\tilde{G}$.
Endpoints in the distributions of invariant masses
$m_{\ell\ell}$, $m_{\ell\gamma}$, 
and $m_{\ell\ell\gamma}$ are combined to solve  $m_{\tilde{\ell}}$,
$m_{\tilde{\chi}^0_1}$ and $m_{\tilde{\chi}^0_2}$.
Note that  only the 
events near the endpoints contribute to the mass determination
in the endpoint analysis.

Our proposal is quite different from the endpoint analysis.  We assume
a set of events come from a common cascade decay, and use the mass
shell condition of the sparticles involved in the cascade decay.  Each
event gives an independent constraint to the masses as given in
Eq.~(\ref{mass}), and contributes to the mass determination. We call
this `` the mass relation method''.  This technique may be applied for
other cascade decays of SUSY particles, which will be discussed in
future publications.

The ATLAS detector at the LHC has a good capability to measure
non-pointing photons, where the barrel part of the electromagnetic
calorimeter (ECAL) and the transition radiation tracker (TRT) will
play important roles (FIG.~\ref{nlsp}).  The barrel ECAL is a
liquid-Argon calorimeter, and covers the psuedo-rapidity range $|\eta|
< 1.4$.  The inner radius of the ECAL is 150~cm.  We assume the
angular resolutions of the photon arrival point at the ECAL inner
surface to be $\sigma_{\phi}\sim 0.004$ and $\sigma_{\eta} \sim
0.002$.  The ECAL energy resolution is expected to be
$\sigma_{E_{\gamma}} / E_{\gamma} = 10\%/\sqrt{E_{\gamma}}$, where the
photon energy $E_{\gamma}$ is given in GeV.  The longitudinal and
transverse segmentation of the ECAL gives a measure of the development
of electromagnetic showers.  The first longitudinal sampling is finely
segmented in the $\eta$ direction, resulting in a good angular
resolution of $\sigma_{\theta} = 60\ {\rm mrad}/\sqrt{E_{\gamma}}$,
where $\theta$ is the polar angle of the photon momentum with respect
to the beam axis.  The azimuthal angle $\phi$ of the photon momentum
is only poorly measured by the ECAL, as the segmentation is very
coarse in this direction.  The ECAL also has an excellent time
resolution, $\sigma_{t_{\gamma}} < 100$~ps for $E_{\gamma} > 30$~GeV,
confirmed by a test-beam experiment.

When a photon is pointing to the interaction point, the photon
momentum is precisely determined by the ECAL only, namely by measuring
the energy deposit and the arrival position.  However, in the case of
the GM models, the photon is in general non-pointing, and the
transverse components of the photon momentum are only poorly measured.
Fortunately, the barrel TRT is located inside the barrel ECAL as a
component of the inner tracking detector.  This detector covers the
radial range from 56 to 107~cm and the pseudo-rapidity range
$|\eta|<0.7$.  As the straw tube trackers of the barrel TRT are
parallel to the beam axis, trajectories of charged particles are
precisely measured in the $r$-$\phi$ plane.  If a photon is converted
into an $e^+e^-$ pair before escaping the barrel TRT, the $\phi$ angle
of the photon momentum can be very precisely measured.  As the
${\phi}$ angle resolution is much better than the $\theta$ angle
resolution by the ECAL, the resolution of the angle $\alpha$ becomes
$\sigma_{\alpha} =
\sqrt{\sigma_{\theta}^2 + \sigma_{\phi}^2} \sim \sigma_{\theta}$.  The
material thickness of the TRT is about 10\% of one radiation length at
$\eta \sim 0$. Namely $\sim$10\% of photons will be converted in the TRT.

\section{Reconstruction of Gravitino direction and Sparticle masses}

In order to test the new techniques we perform a Monte Carlo event
simulation at GM point~G1 \cite{TDR,Hinchliffe:1998ys} with $F_0$, or
equivalently the neutralino lifetime $c\tau$, being the only free
parameter.  The model parameters and some of sparticle masses are
listed in TABLE~\ref{G1}.  The low energy SUSY parameters are
calculated by ISASUSY 7.51, and the mass spectrum, the couplings and
the decay branching ratios are interfaced to HERWIG version 6.4.  We
generate $10^5$ SUSY events at this point, corresponding to an
integrated luminosity of 13.9~fb$^{-1}$.  When we simulate the events,
we keep the lightest neutralinos (NLSPs) stable at the generator
level.  Then a fast detector simulator ATLFAST is used for all
particles except the neutralinos.  The decay of the neutralinos and
the photon conversions are simulated at the analysis stage.  The
photon conversion probability is estimated based on the detector
thickness of the TRT~\cite{TDR}.  If (a) a neutralino decays into a
photon and a gravitino before escaping the TRT region, (b) the photon
points to the barrel ECAL, and (c) the photon is converted inside the
barrel TRT, then the energy, position, time, and direction of the
photon are smeared by the Gaussian distribution according to their
resolutions.  The detector resolutions assumed in the simulation are
listed in TABLE~\ref{reso}.  We assume the time resolution of the ECAL
to be constant $\sigma_{t_{\gamma}} = 100$~ps for $E_{\gamma} >
30$~GeV.

\begin{table}
\caption{\label{G1}
Model parameters 
and some of the sparticle masses at point G1.
The parameter
$N$ is an integer number which appears in Eq.~(\ref{mass1}).
}
\begin{ruledtabular}
\begin{tabular}{cccc}
Parameters & \multicolumn{3}{c}{Sparticle masses (GeV)} \\ \hline
$F/M=90$~TeV & $m_{\tilde{g}}=720$
& $m_{\tilde{\ell}_L}=324$ & $m_{\NLSP}=117$\\
$M=500$~TeV & $m_{\tilde{q}_L}=958$ 
& $m_{\tilde{\ell}_R}=162$ & $m_{\tilde{\chi}_2^0}=217$\\ 
$N=1$ & $m_{\tilde{q}_R}=915$ 
& $m_{\tilde{t}_1}=831$ & $m_{\tilde{\chi}_3^0}=420$ \\
$\tan\beta=5$ & &$m_{\tilde{b}_1}=909$ &  $m_{\tilde{\chi}_4^0}=442$ \\ 
$\mu>0$ & & & \\
\end{tabular}
\end{ruledtabular}
\end{table}

\begin{table}

\caption{\label{reso}
Detector resolutions assumed in our Monte Carlo simulation
of non-pointing photons.
The photon energy $E_{\gamma}$ is given in GeV.
As for the $\phi$ angle of the photon momentum,
we assume the photon conversion in the TRT detector.}
\begin{ruledtabular}
\begin{tabular}{lccc}
\multicolumn{2}{c}{Observable} & Detector & Resolution \\ \hline
Photon energy  & $E_{\gamma}$ & ECAL & $0.1 \sqrt{E_{\gamma}}$ \\ \hline
Photon arrival time & $t_{\gamma}$ & ECAL & 100~ps \\ \hline
Photon arrival position & $\eta$ & ECAL & 0.002 \\
& $\phi$ & ECAL & 0.004 \\ \hline
Photon momentum & $\theta$ 
& ECAL & 0.060/$\sqrt{E_{\gamma}}$ \\
& $\phi$ & TRT & 0.001 \\
\end{tabular}
\end{ruledtabular}
\end{table}

For a moment we set the neutralino lifetime
to be
$c\tau = 100$~cm.
In this case
the neutralinos efficiently decay 
in the TRT,
as the outer radius of the TRT is roughly 100~cm.
We first apply pre-selection cuts 
to suppress the Standard Model background:
\begin{eqnarray}
{\rm i}) && M_{\rm eff} > 400~{\rm GeV}, \nonumber \\
{\rm ii}) && E_{\rm T}^{\rm miss}> 0.1 M_{\rm eff}.
\label{cut0}
\end{eqnarray}
The missing transverse energy $E_{\rm T}^{\rm miss}$ is calculated
from the reconstructed jets, leptons, photons and unreconstructed
calorimeter energies.  The effective mass is defined by the sum of the
missing transverse energy and the transverse momenta of the four
hardest jets:
\begin{equation}
M_{\rm eff} = E_{\rm T}^{\rm miss} 
+ p_{\rm T,1}+p_{\rm T,2}+p_{\rm T,3}+p_{\rm T,4}
\end{equation}
We do not include photons from the neutralino decays
in the above calculation.
The efficiency of the pre-selection
cuts for the generated SUSY events is 80\%.

\begin{figure}[t]
\centerline{
\centerline{\psfig{file=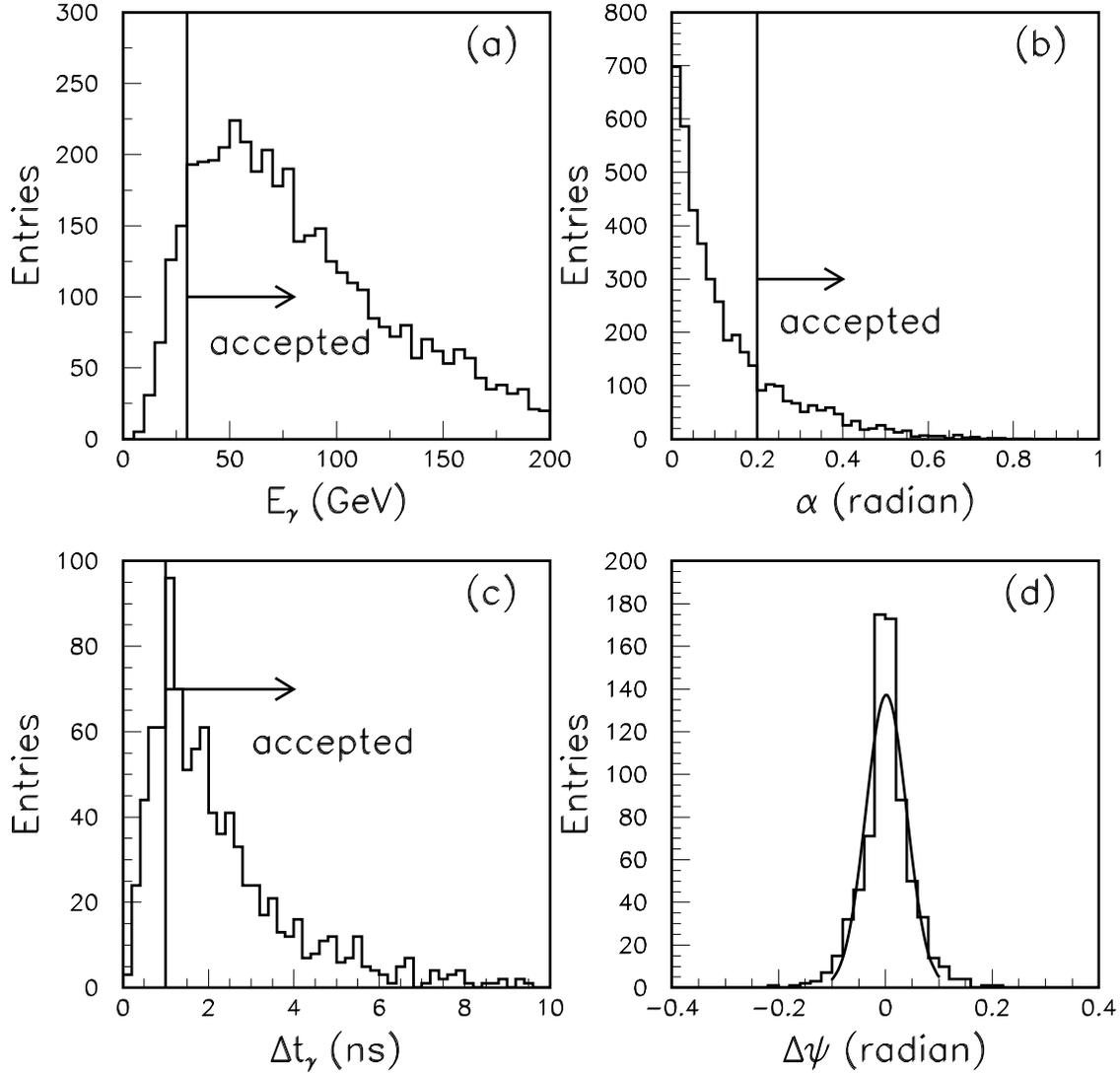,width=\columnwidth}} 
}
\caption{\label{gamsel}
Distributions of 
(a) $E_{\gamma}$, 
(b) $\alpha$,
and (c) $\Delta t_{\gamma}$ 
at point G1 with $c\tau=100$ cm,
where the cuts 1)--3) in Eq.~(\ref{cut1}) are sequentially applied.  
(d) Distribution of 
$\Delta\psi \equiv \psi-\psi_{\rm true}$
after applying the cuts, where $\psi$ is the angle between 
the momenta of the photon and the gravitino. 
The result of a Gaussian fit is also shown.}
\end{figure}

The following cuts\footnote{
Isolation from tracks/clusters is yet to be examined.}
are then applied to select good `non-pointing' photons
with conversion in the TRT:
\begin{eqnarray}
1) && E_{\gamma} > 30~{\rm GeV}, \nonumber \\
2) && \alpha > 0.2, \nonumber \\
3) && \Delta t_{\gamma}
(\equiv t_{\gamma}- L/c) 
> 1.0~{\rm ns}.
\label{cut1}
\end{eqnarray}
The distributions of $E_{\gamma}$, $\alpha$
and $\Delta t_{\gamma}$ are shown  
in FIGs.~\ref{gamsel}(a)--(c),
where the cuts 1)--3) are sequentially applied.

In FIG.~\ref{gamsel}(d) we plot $\Delta\psi \equiv \psi - \psi_{\rm
true}$, where $\psi$ is calculated from the measured $L$, $\alpha$,
and $t_{\gamma}$ using Eq.~(\ref{kobayashi}) and $\psi_{\rm true}$ is
the true value obtained from the generator information.  The
resolution $\sigma_{\psi}$ is better than 40~mrad in this case.

\begin{figure}
\centerline{\psfig{file=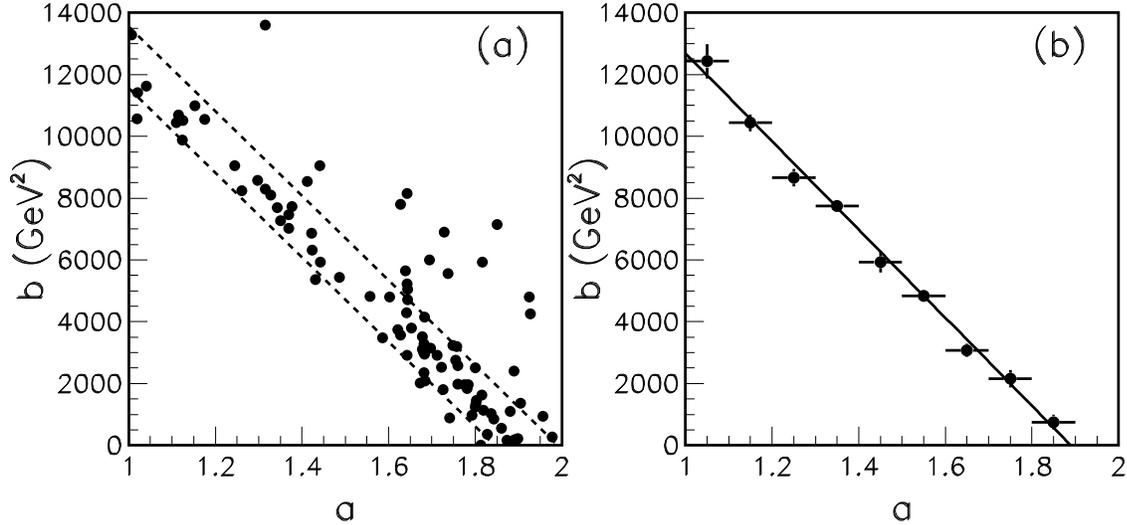,width=\columnwidth}}
\caption{\label{ab}
(a) Distribution of $\ell\gamma$ pairs 
in $(a,b)$ plane. The region between dotted lines 
shows our selection cut.
(b) Points show the average values 
of $(a,b)$ between the two dotted lines in (a). 
The region 
is divided into 9 bins. The solid line shows a result of a linear fit.}
\end{figure}

In order to determine the masses of the slepton and the neutralino
which appear in the cascade decay $\tilde{\ell} \to \ell\NLSP \to
\ell\gamma\tilde{G}$, isolated leptons (electrons and muons) with
transverse momentum larger than 20~GeV are searched for to make a pair
with the non-pointing photon.  If there are several leptons in an
event, we choose the $\ell\gamma$ pair which minimizes the invariant
mass $m_{\ell\gamma}$.  The parameters $a$ and $b$ are calculated for
each $\ell\gamma$ pair.  The scatter plot in the $(a,b)$ plane is
shown in FIG.~\ref{ab}(a), where the sample contains about 120
$\ell\gamma$ pairs.  The points are clearly concentrated along a line.
The section of the $b$ axis and the slope of the line must correspond
to $m_{\tilde{\ell}}^2$ and $m_{\NLSP}^2$ because of the relation $b =
m_{\tilde{\ell}}^2 - m_{\NLSP}^2 a\ . $ In FIG.~\ref{ab}(b) we take a
simple average of $(a, b)$ for the events between the two dotted lines
in FIG.~\ref{ab}(a) by dividing the region of $a$ into 9 bins, and we
fit the averaged data by the linear function Eq.~(\ref{mass}).  The
fit results are $m_{\tilde{\ell}}=162.1$~GeV and
$m_{\NLSP}=117.5$~GeV, while the input values are 161.7~GeV and
117.0~GeV, respectively.

\begin{figure}
\centerline{\psfig{file=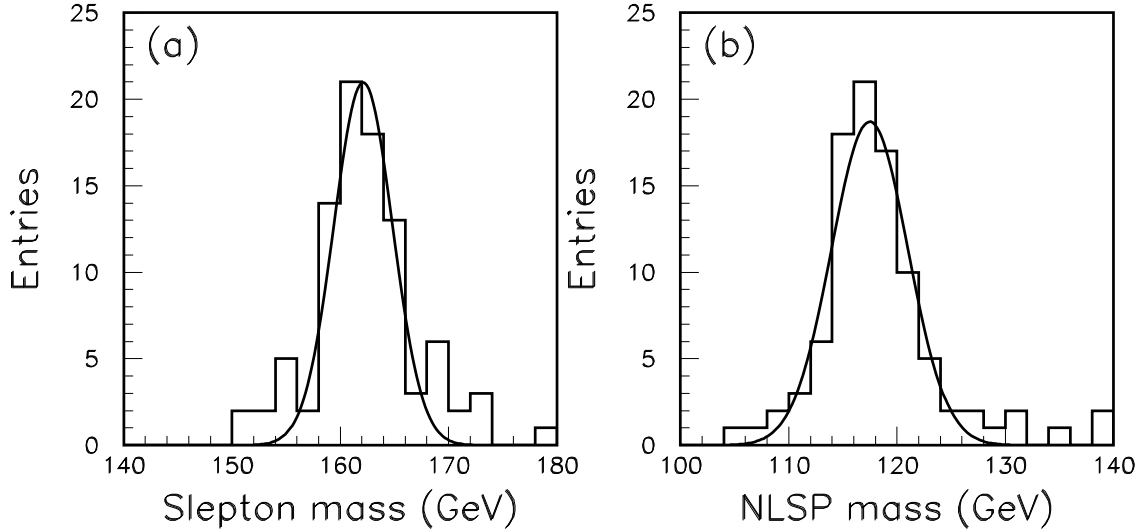,width=\columnwidth}}
\caption{\label{nlspfit}
Distributions of  the fit results of 
(a) the slepton mass $m_{\tilde{\ell}}$
and (b) the neutralino mass $m_{\NLSP}$. 
The simulation and the fit are repeated
100 times with different random number seeds. 
Results of Gaussian fitting are also shown.}
\end{figure}

To estimate errors of the mass measurement, 
the simulation and the fit are repeated
100 times with different random number seeds. 
The fitted masses
are plotted in FIG.~\ref{nlspfit}. 
By fitting the distribution, we obtain the errors of the masses as  
$\sigma_{m_{\tilde{\ell}}}=2.7$~GeV and $\sigma_{m_{\NLSP}}=3.5$~GeV. 
They correspond to relative mass errors of $2 \sim 3$\%. 
If one of the sparticle masses is precisely determined by
some other measurements,
the other sparticle mass can be determined with a precision of $\sim 300$~MeV
by the mass relation method.

\section{Full reconstruction and lifetime measurement}

In this section we demonstrate full reconstruction of 
the cascade decay  
$\tilde{\ell} \rightarrow 
\ell \NLSP  \rightarrow 
\ell \gamma \tilde{G}$,
and we show
that the neutralino lifetime can be determined
by the reconstructed  decay time and momentum of 
the neutralino.
This analysis becomes possible
after the precise determination of $m_{\tilde{\ell}}$ and 
$m_{\NLSP}$ described in the previous section.  

Here we study events with leptons
and non-pointing photons,  
where the photons may or may not be converted 
in the inner detector.
For each $\ell\gamma$ pair, 
the arrival time  $t_{\gamma}$, arrival point 
$\vec{x}_{\gamma}$ and energy $E_{\gamma}$ of the photon, 
the longitudinal component of the photon momentum, 
and the lepton momentum $\vec{p}_{\ell}$ would be directly measured,
while
the transverse components of the photon momentum, 
the neutralino 
four  momentum $p_{\tilde{\chi}}$ and its decay time
$t_{D}$ $(<t_{\gamma})$ 
are not directly measured. On the other hand, 
we have the following equations involving the unknown 
quantities,  
\begin{eqnarray}
\vec{v}_{\tilde{\chi}} t_D + \vec{v}_{\gamma}(t_{\gamma}-t_D)
&=& \vec{x}_{\gamma},\cr
(p_{\tilde{\chi}}+ p_{l})^2&=&m^2_{\tilde{\ell}},\cr
(p_{\tilde{\chi}}-p_{\gamma})^2 &=& p^2_{\tilde{G}}
=m^2_{\tilde{G}} = 0,\cr
p^2_{\tilde{\chi}}&=&m^2_{\tilde{\chi}}\ ,
\label{fulleq}
\end{eqnarray}
where
$\vec{v}_{\tilde{\chi}}$ and $\vec{v}_{\gamma}$
are velocity vectors ($|\vec{v}_{\gamma}|=c$),
and the lepton and gravitino masses are neglected.
Provided that we know the two masses 
$m_{\tilde{\ell}}$ and $m_{\NLSP}$,
we can solve all the unknown parameters from the equations.
There are two solutions for each $\ell\gamma$ pair,   
which we obtain by numerically solving Eq.~(\ref{fulleq}).

The simulation and analysis are modified from 
those of the previous section.
The neutralino must decay inside the ECAL and
the photon from the neutralino decay must enter the barrel ECAL,
resulting in a much larger fiducial decay volume.
We assume the fiducial volume as
$r < 150$~cm and $|z| < 300$~cm,
where $r$ and $z$ are the radial distance from the beam axis
and the distance from the interaction point along the beam axis,
respectively.
In addition, 
as the photon conversion is not required
in the inner detector,
the acceptance of the events further increases. 
After applying the pre-selection cuts
given in Eq.~(\ref{cut0}),
we select non-pointing photons by the following cuts;
\begin{eqnarray}
1) && E_{\gamma}>30~{\rm GeV},\cr
2) && \Delta \theta>0.2~{\rm rad}, \cr
3) && \Delta t_{\gamma}>1~{\rm ns}. 
\label{cuts2}
\end{eqnarray}
Here 
$\Delta \theta \equiv \theta_{\vec{p}_{\gamma}} -
\theta_{\vec{x}_{\gamma}}$ 
is the difference between the 
polar angles of $\vec{p}_{\gamma}$ and $\vec{x}_{\gamma}$,
where the polar angles
are measured from the beam axis. Note that the cut 2) is different 
from that in Eq.~(\ref{cut1}), because one cannot determine 
the angle $\alpha$ without precise measurement of the $\phi$ angle
using the photon conversion.
This $\Delta \theta$ cut is less efficient than the $\alpha$ cut
in Eq.~(\ref{cut1}).
However, the overall selection efficiency of this analysis
is much larger 
because we have the larger fiducial decay volume  
and do not require the photon conversion in the inner detector. 

Isolated leptons with transverse momentum 
larger than 20~GeV
are used to make $\ell\gamma$ pairs.
For each $\ell\gamma$ pair, there are two solutions 
for decay kinematics. 
We take solutions if the reconstructed decay points
are in the fiducial volume (e.g. $r<100$~cm and $|z|<300$~cm).
Therefore each $\ell\gamma$ pair may be counted twice.
If there are several leptons in an event, all leptons are tried for
each non-pointing photon, and accepted if the $\ell\gamma$ pair satisfies  
the fiducial condition. 

\begin{figure}
\centerline{\psfig{file=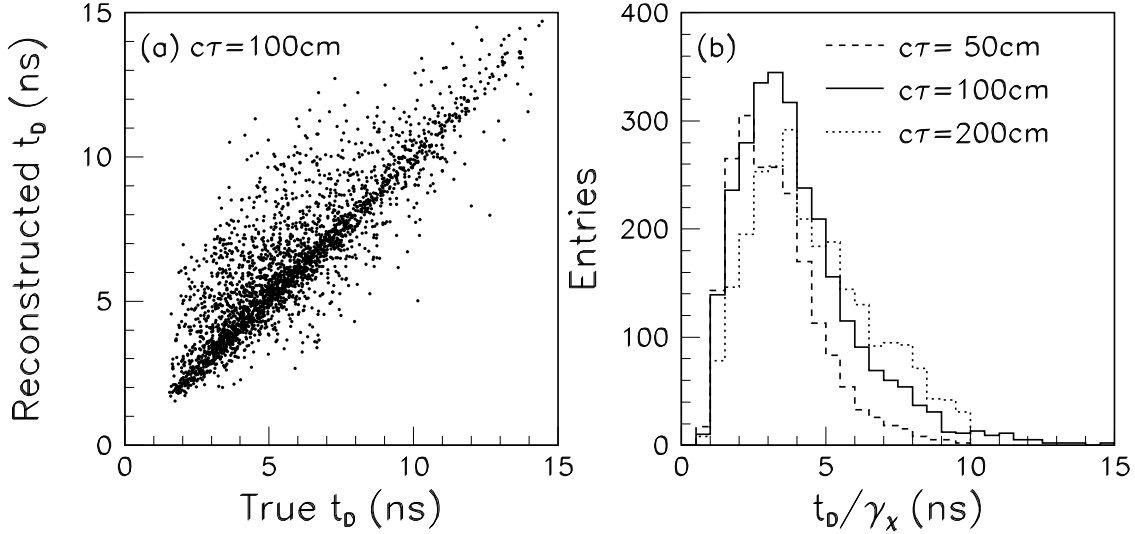,width=\columnwidth}}
\caption{\label{ctauplot}
(a) The distribution of the neutralino decay time $t_D$ vs 
the true value. 
(b) Distributions of $t_D/\gamma_{\chi}$  
for $c\tau=50$~cm (dashed), 100~cm (solid)
and 200~cm (dotted). }
\end{figure}

In FIG.~\ref{ctauplot}(a) we show 
a scatter plot of the reconstructed decay time $t_D$ and 
the true decay time
for $c\tau=100$~cm. Here we use the input $m_{\tilde{\ell}}$ and 
$m_{\tilde{\chi}^0_1}$ for reconstruction. 
This plot shows that the neutralino decay time 
is correctly reconstructed  
for a significant fraction of the events.  
We also show the distribution of 
$t_D/\gamma_{{\chi}}$ in FIG.~\ref{ctauplot}(b) for $c \tau=50$~cm,
100~cm, and 200~cm,
where the factor
$1/\gamma_{{\chi}}$  corrects the effect of 
Lorenz boost.
The distribution 
shows the expected  exponential damping toward large 
$t_D/\gamma_{{\chi}}$ values.
The geometric acceptance of the events would be small when 
$c\tau\gamma_{\chi} > R_{in}$ or
$\tau\gamma_{\chi}<1$~ns, 
where $R_{in}$ ($\sim 150$~cm) is the inner radius of the ECAL and
1~ns is our cut value on the difference of the photon arrival time.  

The geometric effect to the
acceptance may be corrected by a study
with full detector simulations. 
The momentum distribution of the neutralino
would be an important uncertainty, 
as the transverse momentum distribution 
should depend on the gluino and squark masses. 
The heavier sparticle masses may be measured precisely by 
the $j\tilde{\chi}^0_1$  or $jj\tilde{\chi}^0_1$
invariant mass distribution for selected jets ($j$)
and a neutralino, where 
the neutralino momentum is reconstructed by solving Eqs.~(\ref{fulleq}). 
Therefore, 
we assume  the systematics 
due to the momentum distribution of the neutralino 
would be small enough.

\begin{figure}
\centerline{\psfig{file=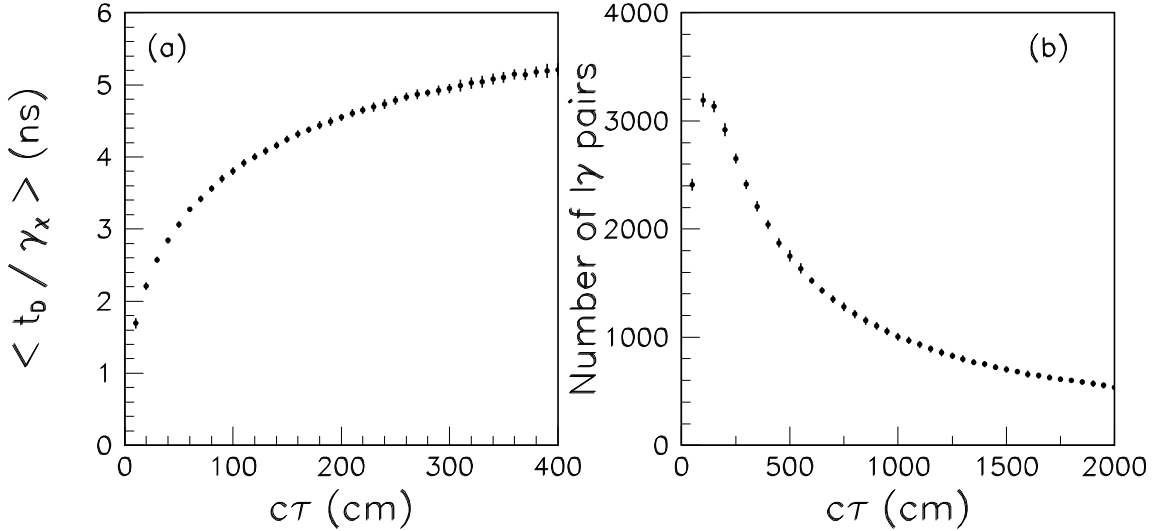,width=\columnwidth}}
\caption{\label{avedtg}
(a) Average $\langle t_D/\gamma_{\chi} \rangle$ and 
(b) $N_{\ell\gamma}$ 
as functions of $c\tau$ 
for an integrated luminosity
of 13.9 fb$^{-1}$ at point G1.  
A dot and an error bar show the mean value and 
the standard deviation of 100 simulations at each $c\tau$ value,
respectively.
}
\end{figure}

Assuming that the systematic errors are controlled, we
may study the sensitivity to the lifetime $c\tau$ 
using the two measured values;
\begin{itemize}
\item[(a)] 
$\langle t_D/\gamma_{{\chi}} \rangle$:
the average of the corrected decay time,
\item[(b)]
$N_{\ell\gamma}$:
the number of accepted $\ell\gamma$ pairs.
Note that we do not count an $\ell\gamma$ pair
more than once to avoid  over-counting.
\end{itemize}
They are plotted as functions of $c\tau$ in FIG.~\ref{avedtg},
where we repeat the
simulation for an integrated luminosity of
13.9~fb$^{-1}$ hundred times to obtain the mean
$\langle t_D/\gamma_{{\chi}} \rangle$ and $N_{\ell\gamma}$ values
and their standard deviations.  
In the plot,
$\langle t_D/\gamma_{{\chi}}\rangle$ 
is larger than 1~ns even for $c\tau=10$~cm
because of the cut $\Delta t_{\gamma} >1$~ns.  
Note that we do not have sensitivity
on $\psi$ when $c\tau<10$~cm because the time resolution is $\sim$0.1~ns,
and $c\times 0.1$~ns $\sim 3$~cm.
The number of $\ell\gamma$ pairs with photon conversion
($N_{\ell\gamma}^{\rm conv}$)
is 13.0 and 80.5 for $c\tau=10$~cm
and $c\tau=30$~cm, respectively.
As the number of converted photons
is rather small for $c\tau=10$~cm,
a large integrated luminosity is needed 
to determine the sparticle masses.
For large $c\tau$ values, the average  $\langle
t_D/\gamma_{{\chi}} \rangle$ is 
saturated since
most of the neutralinos decay outside the
detector. Indeed the number of the reconstructed events takes its maximum
at $c\tau\sim 100$~cm and decreases monotonically as shown
in FIG.~\ref{avedtg}(b).

\begin{figure}
\centerline{\psfig{file=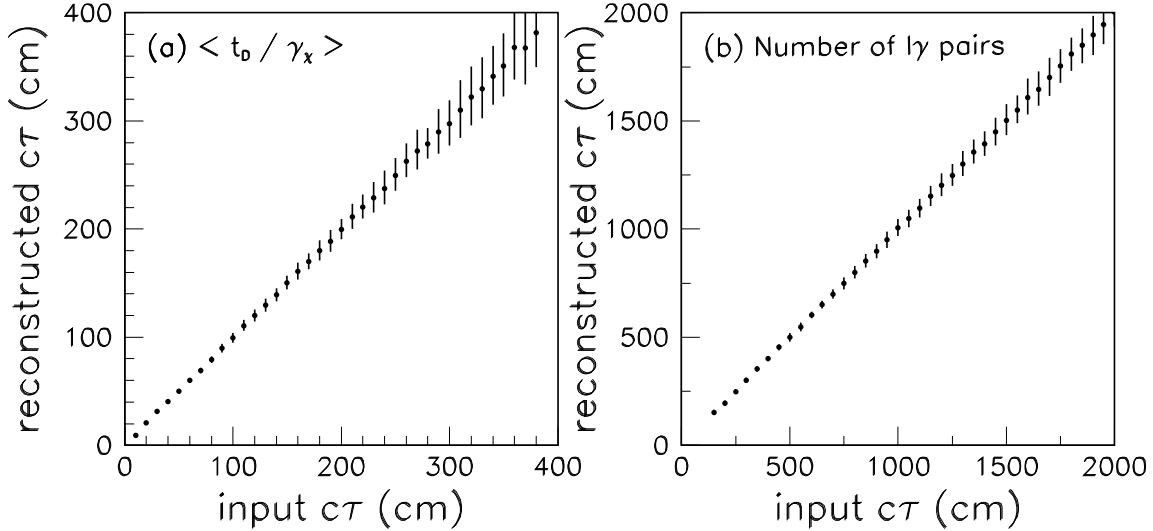,width=\columnwidth}}
\caption{\label{taures}
Estimated resolution of 
the lifetime $c\tau$ for 
an integrated luminosity of 13.9~fb$^{-1}$ from 
(a) the average $t_{D}/\gamma_{\chi}$ and 
(b) the number of $\ell\gamma$ pairs $N_{\ell\gamma}$. The input 
$m_{\tilde{\ell}}$ and $m_{\tilde{\chi}}$ are used for the reconstruction
and their errors are ignored.}
\end{figure}

The sensitivity of the $\langle t_D/\gamma_{{\chi}} \rangle$
measurement 
to the lifetime $c\tau$ is estimated by the error of the
measurement $\Delta \langle t_D/\gamma_{{\chi}} \rangle$.
Namely we define the error of the lifetime $\Delta c\tau$
by the following formula:
\begin{equation}
g(c\tau \pm \Delta c\tau)
= \langle t_D/\gamma_{{\chi}} \rangle
\pm \Delta \langle t_D/\gamma_{{\chi}} \rangle\ , 
\end{equation}
where $g(c\tau)$ is a function to describe  
the average $\langle t_D/\gamma_{\chi} \rangle$.
This function is numerically obtained by 
fitting 
the $\langle t_D/\gamma_{{\chi}} \rangle$ to
a second power polynomial function of $c\tau$
using the average values in the region within $c\tau \pm 50$~cm.
When $c\tau\gg 100$~cm, most of the neutralinos 
decay after reaching the ECAL
and $\langle t_D/\gamma_{\chi}\rangle$
looses the sensitivity
to $c\tau$.  
On the other hand, the number of the $\ell\gamma$ pairs
$N_{\ell\gamma}$ 
sensitively reduces when $c\tau$ is longer than 100~cm. 
We can estimate the sensitivity of $N_{\ell\gamma}$ to $c\tau$
in a same way. 
The result is shown in
FIG~.\ref{taures} and summarized in TABLE~\ref{table1}.
Note that we ignore the errors of $m_{\tilde{\ell}}$ and 
$m_{\NLSP}$
for a moment and use the input mass values.  
Note also that only statistical errors are considered here.
Systematic errors are yet to be studied
with full detector simulations.

\begin{table}
\caption{\label{table1}
Relative resolution of $c\tau$.
As in FIG.~\ref{taures}
(a) and (b) are estimated from the average 
$\langle t_{D}/\gamma_{\chi} \rangle$ and 
the number of $\ell\gamma$ pairs $N_{\gamma\ell}$, respectively.
}
\begin{ruledtabular}
\begin{tabular}{rrrr}
\multicolumn{2}{c}{(a) From $\langle t_D/\gamma_{\chi} \rangle$}
& \multicolumn{2}{c}{(b) From $N_{\ell\gamma}$} \\
$c\tau$~(cm) & $\Delta c\tau/c\tau$ & $c\tau$(cm) 
& $\Delta c\tau/c\tau$\cr
\hline
10 & 0.173 & 200 & 0.062 \cr
20 & 0.062 & 250 & 0.038 \cr
30 & 0.041 & 300 & 0.034 \cr
40 & 0.036 & 350 & 0.036 \cr
50 & 0.037 & 400 & 0.030 \cr
60 & 0.027 & 450 & 0.031 \cr
70 & 0.037 & 500 & 0.040 \cr
80 & 0.039 & 600 & 0.025 \cr
90 & 0.043 &   700 & 0.037 \cr
100 & 0.045 &  800 & 0.037 \cr
150 & 0.043 &  900 & 0.037 \cr
200 & 0.047 & 1000 & 0.039 \cr
250 & 0.061 & 1500 & 0.048 \cr
300 & 0.070 & 2000 & 0.033 \cr
\end{tabular}
\end{ruledtabular}
\end{table}

For large $c\tau$ values 
the number of selected events 
is limited.
In TABLE~\ref{table2} we summarize $N_{\ell\gamma}$
and $N_{\ell\gamma}^{\rm conv}$ 
for an integrated luminosity of  13.9~fb$^{-1}$.
For $c\tau<2000$~cm the number $N_{\ell\gamma}^{\rm conv}$
still exceeds 130
events if we assume an integrated luminosity of 
100~fb$^{-1}$ at point G1.  
Therefore, it might be still possible to determine the neutralino
and slepton masses.

\begin{table}
\caption{\label{table2}
Number of $\ell\gamma$ pairs 
for an integrated luminosity of 13.9~fb$^{-1}$.
$N_{\ell\gamma}^{\rm conv}$ is  the number of  $\ell\gamma$ pairs 
with photon is conversion in the inner detector.  
}
\begin{ruledtabular}
\begin{tabular}{rrr}
$c\tau$~(cm)& $N_{\ell\gamma}$ & $N_{\ell\gamma}^{\rm conv}$  \cr
\hline
10 & 147 & 13.0 \cr
30 & 1424 & 80.5 \cr
100 & 3193 & 116.3 \cr
300 & 2413 &  72.8\cr
2000& 536 & 14.3\cr
\end{tabular}
\end{ruledtabular}
\end{table}

\section{Determination of the fundamental parameters}

In this section
we utilize the measurement
of the sparticle masses 
and the neutralino lifetime 
to determine more fundamental parameters in the GM scenario.
From the measurement of the   
slepton and neutralino masses, 
the ratio $F/M$ is determined
with a precision of a few \%.
This can be seen in the expression of the masses at the messenger
scale as~\cite{Martin:1996zb};
\begin{eqnarray}
M_i&=&  \frac{\alpha_i(M)}{4\pi} 
\frac{F}{M} N \times g\left( F/M^2\right)
\cr
m^2_{\tilde{\ell}_R} &=&\frac{3\alpha_1^2(M)}{40\pi^2} 
\frac{F^2}{M^2} N \times f\left(F/M^2\right)\cr
m^2_{\tilde{\ell}_L} &=& 
\left( \frac{3\alpha_1^2(M)}{160\pi^2} 
+\frac{3\alpha_2^2(M)}{32\pi^2} \right) 
\frac{F^2}{M^2} N \times f\left(F/M^2\right)\ , 
\label{mass1}
\end{eqnarray}
where $N$ is an integer number, while
$g$ and $f$ are some functions which satisfy
$f(0)=g(0)\sim 1$ for $F\ll M^2$. The sparticle masses are
proportional to $F/M$ in this limit.  

The absolute size of $M$ (or $F$) is rather difficult to
determine because it only appears through the sfermion mass running from
$M$ to the SUSY scale. A study~\cite{TDR} shows that 
the relative error $\Delta M/M$ is $\sim 30$\%
with 1\% mass errors at this point.

The neutralino lifetime
depends on the order parameter of the total SUSY breaking $F_0$ and
the neutralino mass $m_{\NLSP}$~\cite{Fayet:vd};  
\begin{equation}
c \tau= \frac{1}{k_{\gamma}}
\left(\frac{100\ \GeV}{m_{\NLSP}}\right)^5
\left(\frac{\sqrt{F_0}}{100\ {\rm TeV}}\right)^4\times 10^{-2}\ {\rm cm}\ ,
\label{life}
\end{equation}
where $k_{\gamma}=\vert N_{11}\cos\theta_W + N_{12}\sin\theta_W\vert^2$
with $\theta_W$ being the Weinberg angle,
and $N_{ij}$ is the neutralino mixing angles. 
The constant is $k_{\gamma}=\cos^2\theta_W$
for the bino-like $\tilde{\chi}^0_1$. 
The parameter $F_0$ is also related to the
gravitino mass itself~\cite{Volkov:jd,Deser:uq};
\begin{equation}
m_{\tilde{G}}=\frac{1}{\sqrt{3}}
\frac{F_0}{M_{pl}}=\left(\frac{\sqrt{F_0}}{100\ \TeV}\right)^2 
2.4\ {\rm eV}.
\label{mass2}
\end{equation}

The sensitivity to the neutralino lifetime
is about 4\% for $c\tau\sim 100$~cm as given in the previous
section. Ignoring the errors of the neutralino  mass and
$k_{\gamma}$, this is translated to 
an error of about 1\% for $\sqrt{F_0}$. 
The lifetime $c\tau=100$~cm corresponds to 
$\sqrt{F_0}\sim 1000$~TeV and
$m_{\tilde{G}} \sim 0.2$~keV, respectively.

If $c\tau$ is ${\cal O}(10) \sim {\cal O}(100)$~m,
the ratio of the number of $\ell\gamma$ pairs
to that of SUGRA-like events is a more sensitive measure for the lifetime.
For $c\tau=2000$~cm ($\sqrt{F_0}\sim2400$~TeV and
$m_{\tilde{G}}\sim 1.4$~keV), the estimated error is about 4\%. 
This corresponds to a 1(2)\% error on $\sqrt{F_0}$ ($m_{\tilde{G}}$).
The mass error of $\tilde{\chi}^0_1$ is estimated to be
around 3\%, because the number of events with a converted photon is 
around 100 for an integrated luminosity of 100~fb$^{-1}$,
which is obtained in one year at the high luminosity runs of the LHC. 
From Eqs.~(\ref{mass1}) and (\ref{mass2})
the error of $F_0$ ($m_{\tilde{G}}$)
due to the mass error is 3(6)\% in this case. 
For $c\tau=10^4$~cm 
the number of events with a lepton and a converted photon is
about 70  for the ultimate integrated luminosity of
300~fb$^{-1}$.  

The above estimate is rather optimistic because
we do not consider systematic errors of the measurement.
In addition, 
we have so far not considered the effect of background events
from the Standard Model processes and the SUSY production itself,
where prompt photons from the interaction point
mimic non-pointing photons due to the limited detector resolution.
Note that in our Monte Carlo sample, 
28757 sleptons are
produced, and  $\sim 2500$ ($\sim$70) 
of them are accepted as the sample 
with a (converted) non-pointing photon 
for $c\tau=300$~cm.  
Especially for a large $c\tau$ case,
the number of real non-pointing photons 
is small and the analysis might severely suffer from the background.
Studies with full detector simulations are necessary
to understand the systematic errors and the background,
and should be completed before the start of the LHC physics run.
If events with a non-pointing photon
converting in the TRT 
cannot be used for mass reconstruction due 
to the backgrounds, the $\Delta
m_{\tilde{\chi}^0_1}\sim $ 30 \% is expected from the endpoint analysis
involving jets and leptons.  In this case
the uncertainty of the neutralino mass
would lead to an error of the gravitino mass by a factor of two. 

\begin{acknowledgments}
We thank the ATLAS collaboration members for useful discussion. We
have made use of the physics analysis framework and tools which
are the result of collaboration-wide efforts.
We especially  thank Dr. J.~Kanzaki and D.~Toya. 
We also thank Dr. G.~Polesello for
useful suggestions. 
This work is
supported in part by the Grant-in-Aid for Science Research, Ministry
of Education, Science and Culture, Japan 
(No.11207101 and 15340076 for K.K., and 
No.14540260 and 14046210 for M.M.N.). M.M.N. is also supported in part 
by a Grant-in-Aid for the 21st Cerntury COE ``Cernter for 
Diversity and Universality in Physics.
\end{acknowledgments}

\end{document}